# A Bayesian Compressive Sensing Approach to Robust Near-Field Antenna Characterization

M. Salucci,[(1)]*Member, IEEE*, N. Anselmi,[(1)]*Member, IEEE*, M. D. Migliore,[(2)]*Senior Member*, and A. Massa,[(1)(3)(4)] *Fellow, IEEE*

[(1)] *CNIT* - "University of Trento" Research Unit

Via Sommarive 9, 38123 Trento - Italy

E-mail: {*marco.salucci, nicola.anselmi.1, andrea.massa*}@*unitn.it*

Website: *www.eledia.org/eledia-unitn*

[(2)] *ELEDIA Research Center* (*ELEDIA@UniCAS* - DIEI, University of Cassino and Southern Lazio)

Via Di Biasio 43, I-03043 Cassino, Italy

E-mail: *donald.migliore@eledia.org*

Website: *www.eledia.org/eledia-unicas*

[(3)] *ELEDIA Research Center* (*ELEDIA@UESTC* - UESTC)

School of Electronic Engineering, Chengdu 611731 - China

E-mail: *andrea.massa@uestc.edu.cn*

Website: *www.eledia.org/eledia-uestc*

[(4)] *ELEDIA Research Center* (*ELEDIA@TSINGHUA* - Tsinghua University)

30 Shuangqing Rd, 100084 Haidian, Beijing - China

E-mail: *andrea.massa@tsinghua.edu.cn*

Website: *www.eledia.org/eledia-tsinghua*

# A Bayesian Compressive Sensing Approach to Robust Near-Field Antenna Characterization

M. Salucci, N. Anselmi, M. D. Migliore, and A. Massa

**Abstract**

A novel probabilistic sparsity-promoting method for robust near-field (*NF*) antenna characterization is proposed. It leverages on the measurements-by-design (*MebD*) paradigm and it exploits some *a-priori* information on the antenna under test (*AUT*) to generate an over-complete representation basis. Accordingly, the problem at hand is reformulated in a compressive sensing (*CS*) framework as the retrieval of a maximally-sparse distribution (with respect to the overcomplete basis) from a reduced set of measured data and then it is solved by means of a Bayesian strategy. Representative numerical results are presented to, also comparatively, assess the effectiveness of the proposed approach in reducing the 'burden/cost' of the acquisition process as well as to mitigate (possible) truncation errors when dealing with space-constrained probing systems.

# 1 Introduction

We are nowadays witnessing an extraordinary technological advancement in phased array technology as a key-asset to the forthcoming *5G* and *6G* communications standards [1][2]. High-performance multi-input/multi-output (*MIMO*), cognitive, and multi-beam architectures will undergo mass production to allow an ubiquitous implementation of the internet-of-things (*IoT*)-based next-generation wireless environments [3]. For a fast and reliable antenna certification at the end of large-scale manufacturing processes [4][5], over-the-air measurements clearly constitute the most time/cost-effective option. In such a framework, far-field (*FF*) techniques are one viable and consolidated approach. However, they intrinsically suffer from limitations imposed by outdoor sites such as the vulnerability to weather conditions as well as to reflections from (uncontrollable) environmental obstacles/scatterers. Otherwise, near-field (*NF*) probing methods are alternative solutions that guarantee a higher accuracy and repeatability thanks to the exploitation of fully-controlled indoor environments and the availability of efficient *NF-FF* transformation strategies [6]-[18], even though they are prone to the so-called "truncation error" caused by the limited extension of the scanning surface (e.g., planar [14][17], cylindrical [11], spherical [8], conical [12]) in real anechoic chambers. Moreover, the *IEEE* recommended practice for *NF* measurements [19] states that a reliable assessment of the radiation features of an antenna under test (*AUT*) needs a dense probing step $\Delta \rho$ (i.e., $\Delta \rho \leq \frac{\lambda}{2}$, $\lambda$ being the free-space wavelength). This results in time-consuming acquisition procedures due to the huge number of scanning positions [7]. An effective recipe to avoid/mitigate such issues is to exploit the available *a-priori* information on the *AUT*. As a matter of fact, several methodologies are based on the representation of the radiation behavior of the *AUT* in terms of a set of known basis functions defined with accurate full-wave (*FW*) simulations of the *CAD* models of the antenna, which are typically available from previous stages of the design process [7][8]. Accordingly, the *NF* recovery problem at hand is then re-formulated as the retrieval, from a reduced set of measured data, of the expansion coefficients by means of suitable matching strategies [7] or machine learning tools [8]. Within this line of reasoning, the measurements-by-design (*MebD*) paradigm has been proposed as an effective tool to predict the *AUT* features by exploiting, unlike above-mentioned state-of-the art strategies, the generation of an *over-complete* basis rather than a minimum-redundancy one [20]-[22]. Thanks to this, it is possible to recast the problem at hand as a sparsity-retrieval one suitable for a fruitful exploitation of the Compressive Sensing (*CS*) [20][23]. As for this latter, it is worth pointing out that a reliable application of standard (deterministic) *CS* solvers requires a preliminary check of the restricted isometry property (*RIP*) of the observation operator, which rapidly becomes computationally unaffordable even for small/medium-scale problems [23]. To overcome this issue, the *MebD* is mathematically reformulated in this communication within a probabilistic sparsity-promoting framework and then solved by means of a customized Bayesian *CS* (*BCS*) strategy that avoids cumbersome assessments of the *RIP* compliancy [24]. To the best of the authors' knowledge, the main novelties of this research work lie in: (*i*) the formulation of the *NF* prediction problem within a highly-flexible Bayesian framework dependent on neither the knowledge of the nominal/gold antenna nor on a particular topology of the probing setup; (*ii*) a suitable customization of the *BCS* to yield a robust and reliable solution of the *NF* field estimation in a wide range of applicative scenarios.

The outline of this communication is as follows. The mathematical formulation of the *NF* antenna characterization problem and its *BCS*-based solution method are detailed in Sect. II. Representative results are shown and discussed in Sect. III to assess the effectiveness and the potentialities of the proposed approach. Finally (Sect. IV) some conclusions are drawn.

## 2 Mathematical Formulation

To faithfully retrieve the *FF* pattern features of an *AUT* by means of *NF-FF* transformation rules [18], the radiated tangential electric field distribution must be accurately estimated over a sufficiently large surface $\Psi$ (Fig. 1) to limit as much as possible the so-called "truncation error" [16]. Generally speaking, $\Psi$ is sampled according to the Nyquist's rule by choosing $\Delta \rho = \frac{\lambda}{2}$ ($\Delta \rho$ being the sampling rate along a generic direction of the surface $\Psi$) [19] for yielding the set of $T$ locations $\mathcal{T} = \{\mathbf{r}_t \in \Psi; t = 1, ..., T\}$. Let us express the *NF* distribution in $\mathcal{T}$, $\underline{\widetilde{E}} = \{\widetilde{E}(\mathbf{r}_t); t = 1, ..., T\}$, as the linear combination of $B$ properly-built basis vectors, $\underline{\mathcal{A}} = \{\underline{\mathcal{A}}_b; b = 1, ..., B\}$ ($\underline{\mathcal{A}}_b = \{\mathcal{A}_b(\mathbf{r}_t); t = 1, ..., T\}$ being the $b$-th ($b = 1, ..., B$) one) through a set of unknown coefficients $\underline{w} = \{w_b \in \mathbb{C}; b = 1, ..., B\}$

$$\widetilde{\underline{E}} = \underline{\underline{A}}\,\underline{w}. \qquad (1)$$

According to the *MebD* paradigm [20], the basis $\underline{\underline{A}}$ is defined by exploiting the *a-priori* information on the *AUT* through the following procedure:

1. *Uncertainty Identification*: Identify the set of $C$ uncertainty factors that can cause a deviation of the *AUT* radiation features from the ideal/gold ones (e.g., defects of the beamforming network, manufacturing tolerances, etc ...). For each $c$-th ($c = 1, ..., C$) uncertainty descriptor, $\chi_c$, define a suitable (physically-admissible) variation range $\left[\chi_c^{\min}, \chi_c^{\max}\right]$. Finally, let $c = 1$ and $b = 0$, and go to *Step 2*;

2. *Over-Complete Basis Generation Loop* ($c = 1, ..., C$):

   (a) Uniformly sample the $c$-th descriptor to form the set of $K_c$ configurations $\underline{\chi}_c = \left\{\chi_c^{(k)}; k = 1, ..., K_c\right\}$, the $k$-th one being

   $$\chi_c^{(k)} = \chi_c^{\min} + (k-1)\frac{\left(\chi_c^{\max} - \chi_c^{\min}\right)}{(K_c - 1)}; \qquad (2)$$

   (b) Run $K_c$ *FW* simulations of the *AUT* to fill the set of *NF* distributions $\underline{\underline{\mathcal{E}}}_c = \left\{\underline{\mathcal{E}}_c^{(k)}; k = 1, ..., K_c\right\}$, $\underline{\mathcal{E}}_c^{(k)} = \left\{E\left(\mathbf{r}_t| \chi_c^{(k)}\right); t = 1, ..., T\right\}$ being the sampled *NF* distribution in $\Psi$ for an *AUT* whose $c$-th uncertainty descriptor has a value equal to the $k$-th sample of its variation range (i.e., $\chi_c = \chi_c^{(k)}$);

   (c) Apply the truncated singular value decomposition (*TSVD*) to $\underline{\underline{\mathcal{E}}}_c$

   $$\underline{\underline{\mathcal{E}}}_c = \underline{\underline{\mathcal{U}}}_c \underline{\underline{\Sigma}}_c \left(\underline{\underline{\mathcal{V}}}_c\right)^* \qquad (3)$$

   where $\underline{\underline{\mathcal{U}}}_c = \left\{\underline{\mathcal{U}}_c^{(q)}; q = 1, ..., Q_c\right\}$ and $\underline{\underline{\mathcal{V}}}_c = \left\{\underline{\mathcal{V}}_c^{(q)}; q = 1, ..., Q_c\right\}$ are the first left and right singular vectors associated to the $Q_c$ singular values, $\underline{\sigma}_c = \left\{\sigma_c^{(q)}; q = 1, ..., Q_c\right\}$, above the noise threshold, respectively, while $\underline{\underline{\Sigma}}_c = \mathrm{diag}\left\{\underline{\sigma}_c\right\}$ and $(.)^*$ stands for the complex conjugate;

(d) Add the $Q_c$ left singular vectors to the basis $\underline{\underline{A}}$ by letting

$$\underline{A}_{(b+q)} \leftarrow \underline{\mathcal{U}}_c^{(q)}; \quad q = 1, ..., Q_c \qquad (4)$$

then update the index $b$ [$b \leftarrow (b + Q_c)$];

(e) If $c = C$ then terminate the iterative loop and output the set of $B$ ($B = \sum_{c=1}^{C} Q_c$) basis, $\underline{\underline{A}}$. Otherwise, let $c \leftarrow (c + 1)$ and go to Step 2(a).

As for the unknown coefficient vector $\underline{w}$ in (1), it is retrieved from a limited set of *NF* data collected over the measurement surface $\Psi_0 \subseteq \Psi$ (Fig. 1). Towards this end, let us assume that the field radiated by the *AUT* is measured over a subset of $M \ll T$ probing locations $\mathcal{M} = \{\mathbf{r}_m \in \Psi_0; m = 1, ..., M\}$ ($\mathcal{M} \subset \mathcal{T}$) to collect the data vector $\underline{d} = \{E(\mathbf{r}_m); m = 1, ..., M\}$. Accordingly, $\underline{w}$ can be computed by solving the following system of equations

$$\underline{\underline{A}}' \underline{w} - \underline{d} = \underline{g} \qquad (5)$$

where the observation operator $\underline{\underline{A}}' = \{\underline{A}'_b; b = 1, ..., B\}$ is derived from $\underline{\underline{A}}$ by setting each $b$-th ($b = 1, ..., B$) column to $\underline{A}'_b = \{A'_b(\mathbf{r}_m); m = 1, ..., M\} = \Gamma\{\underline{A}_b\}$, $\Gamma\{.\}$ being an operator extracting the $M$ entries of $\underline{A}_b$ associated to the probing positions of $\mathcal{M}$. Moreover, the vector $\underline{g} = \{g(\mathbf{r}_m); m = 1, ..., M\}$ in (5) models the presence of an additive noise blurring the data. It is worth pointing out that in real scenarios only a very limited subset (or none) of the $C$ uncertainties affects the measured *AUT*. Therefore, the solution vector $\underline{w}$ is generally intrinsically sparse (i.e., $\|\underline{w}\|_0 \ll B$, $\|.\|_0$ being the $\ell_0$-norm) and effective solutions of (5) can be obtained within the *CS* framework by exploiting a probabilistic Bayesian approach [24] to avoid the *RIP* check. More in detail, the linear system of equations (5) is first re-arranged in the following form

$$\underline{\underline{A}} \underline{\omega} - \underline{\delta} = \underline{\gamma} \qquad (6)$$

where

$$\underline{\underline{A}} = \begin{bmatrix} \Re\{\underline{\underline{A}}'\} & -\Im\{\underline{\underline{A}}'\} \\ \Im\{\underline{\underline{A}}'\} & \Re\{\underline{\underline{A}}'\} \end{bmatrix} \qquad (7)$$

$\Re\{.\}$ and $\Im\{.\}$ being the real and imaginary part, respectively, $\underline{\omega} \triangleq [\Re\{\underline{w}\}, \Im\{\underline{w}\}]$, $\underline{\delta} \triangleq [\Re\{\underline{d}\}, \Im\{\underline{d}\}]$, and $\underline{\gamma} \triangleq [\Re\{\underline{g}\}, \Im\{\underline{g}\}]$. Accordingly, the sparsest guess of $\underline{\omega}$ ($\underline{\widetilde{\omega}} \triangleq \{\widetilde{\omega}_b; b = 1, ..., (2 \times B)\}$) that maximizes the *a-posteriori* probability $\mathcal{P}(\underline{\omega}|\underline{\delta})$ is computed as follows [24]

$$\underline{\widetilde{\omega}} = \frac{1}{\widetilde{\eta}} \left[ \frac{(\underline{\underline{A}})^* \underline{\underline{A}}}{\widetilde{\eta}} + \mathrm{diag}\{\underline{\widetilde{\tau}}\} \right]^{-1} (\underline{\underline{A}})^* \underline{\delta}. \quad (8)$$

In (8), $\widetilde{\eta}$ and $\underline{\widetilde{\tau}}$ ($\underline{\widetilde{\tau}} = \{\widetilde{\tau}_b; b = 1, ..., (2 \times B)\}$) are the estimated *BCS* noise variance and the *BCS* hyper-parameters, respectively. They are determined with a fast relevant vector machine (*RVM*)-based local search strategy [23] by maximizing the following likelihood function

$$\Phi(\eta, \underline{\tau}) = \left\{ -\frac{1}{2} \left[ 2M \log 2\pi + \log |\underline{\underline{\Omega}}| + (\underline{\delta})^* (\underline{\underline{\Omega}})^{-1} \underline{\delta} \right] \right\} \quad (9)$$

starting from initial guess of the *BCS* noise variance, $\eta_0$. In (9), $|.|$ is the matrix determinant, $\underline{\underline{\Omega}} = \eta \underline{\underline{I}} + \underline{\underline{A}} (\mathrm{diag}\{\underline{\tau}\})^{-1} (\underline{\underline{A}})^*$, and $\underline{\underline{I}}$ is the identity matrix. Finally, the solution of (5) is computed by re-arranging the entries of the *BCS* vector (8) into the complex-valued expansion weights as follows

$$\underline{\widetilde{w}} = \{(\widetilde{\omega}_b + j\widetilde{\omega}_{b+B}); b = 1, ...., B\} \quad (10)$$

$j = \sqrt{-1}$ being the imaginary unit, while the corresponding estimated field radiated by the *AUT* at the prediction locations $\mathcal{T}$ is retrieved by inputting (10) into (1).

## 3 Numerical Validation

The objective of this Section is two-fold. On the one hand, representative results from an exhaustive numerical study are reported to assess the effectiveness of the proposed *BCS*-based approach also in comparison with a previously-published state-of-the-art *CS* approach based on the orthogonal matching pursuit (*OMP*) [20]. On the other hand, the Section is aimed at providing the interested readers/users with some useful guidelines for its optimal application. As for the results, besides a pictorial representation of the *NF* field reconstructions, the *NF* integral error

$$\Xi \triangleq \frac{\sum_{t=1}^{T} \left| E(\mathbf{r}_t) - \widetilde{E}(\mathbf{r}_t) \right|^2}{\sum_{t=1}^{T} |E(\mathbf{r}_t)|^2} \quad (11)$$

is computed to provide a quantitative index of the "solution quality".

In the benchmark scenario, the reference *AUT* is a linearly-polarized (i.e., $E(\mathbf{r}) = E_x(\mathbf{r})\widehat{\mathbf{x}}$) planar phased array arranged on the $(x, y)$ plane and it is composed by $N = (N_x \times N_y) = (6 \times 10) = 60$ rectangular probe-fed microstrip patches working at $f = 3.6$ [GHz]. The radiators have dimensions $(l_x, l_y) = (2.2 \times 10^{-1}, 3.3 \times 10^{-1})$ [$\lambda$] and they are etched in a $\frac{\lambda}{2}$-spaced square lattice on a dielectric substrate with relative permittivity $\varepsilon_r = 4.7$, loss tangent $\tan \delta = 1.4 \times 10^{-2}$, and thickness $h = 1.9 \times 10^{-2}$ [$\lambda$]. Moreover, the feeding architecture consists of $S = N_y = 10$ uniformly-excited clusters/planks corresponding to the rows of the array, the nominal/gold excitations being set to $\zeta_s = 1.0$ ($s = 1, ..., S$). As for the *NF* set-up, $\Psi_0$ is a square plane of side $L_{\Psi_0} = 20$ [$\lambda$] placed $H = 7$ [$\lambda$] above the *AUT* top surface. As for the probing location set $\mathcal{M}$, it consists of $M = (M_x \times M_y) = (5 \times 5) = 25$ positions uniformly-distributed over $\Psi_0$ with a step of $\Delta_x^{\mathcal{M}} = \Delta_y^{\mathcal{M}} = 5$ [$\lambda$] [20]. To account for mutual coupling effects, the basis $\underline{A}$ has been built by modeling the *AUT* within the Altair FEKO *FW* simulation environment [25] by considering a $\frac{\lambda}{2}$-sampled (i.e., $\Delta x = \Delta y = \Delta \rho$) prediction surface $\Psi = \Psi_0$ ($T = (T_x \times T_y) = (41 \times 41) = 1681 \Rightarrow \frac{M}{T} \approx 1.5\%$). For the numerical study, the antenna at hand is assumed to be potentially affected by $C = (2 \times S) = 20$ deviations from the gold one, which are associated to non-idealities on both the magnitude ($\chi_c = |\zeta_c|$, $[\chi_c^{\min}, \chi_c^{\max}] = [0, 1]$, $c = 1, ..., S$) and the phase ($\chi_c = \angle \zeta_c$, $[\chi_c^{\min}, \chi_c^{\max}] = [-\pi, \pi]$, $c = S + 1, ..., C$) of the excitations of each $s$-th ($s = 1, ..., S$) sub-array. According to the *MebD* guidelines [20], $K_c = 7$ simulations have been performed for each $c$-th ($c = 1, ..., C$) factor to yield a basis of $B = 40$ vectors, $Q_c = 2$ being the number of truncated singular values for each $c$-th index.

A preliminary calibration of the *BCS* setup has been carried out by assuming as reference an *AUT* affected by a partial failure on the excitation coefficient of the $s = 3$-rd row (i.e., $|\zeta_3| = 0.45$ and $\angle \zeta_3 = \frac{\pi}{3}$ [rad]). Towards this end, the initial guess of the *BCS* noise variance, $\eta_0$, for the *RVM*-based maximization of (9) has been varied within the range $10^{-7} \leq \eta_0 \leq 10$ and the value of the *NF* integral error (11) has been computed for different noise levels ($SNR \in [20, 50]$ [dB]). The outcomes of this analysis are summarized in Fig. 2. As expected, the optimal value of $\eta_0$ depends on the *SNR* (e.g., $\eta_0^{opt}|_{SNR=50\,[dB]} < \eta_0^{opt}|_{SNR=40\,[dB]} < \eta_0^{opt}|_{SNR=30\,[dB]} < \eta_0^{opt}|_{SNR=20\,[dB]}$, being $\eta_0^{opt}|_{SNR} = \arg[\min_{\eta_0}(\Xi(\eta_0)|_{SNR})])$ since, by definition, the larger is the noise variances, the lower is the *SNR*. On the other hand, no *a-priori* accurate information on the noise level is available in several practical cases, thus an optimal trade-off value has been chosen as $\eta_0^{opt} \triangleq \frac{\int \eta_0^{opt}|_{SNR} dSNR}{\int dSNR}$ and it has been set here to $\eta_0^{opt} = 10^{-2}$ (Fig. 2). To assess the reliability of such a calibration in general operative conditions, the normalized error map (i.e., $\left|\Delta \widetilde{E}(\mathbf{r}_t)\right| \triangleq \left|\widetilde{E}(\mathbf{r}_t) - E(\mathbf{r}_t)\right| / \max_{\mathbf{r}_t} |E(\mathbf{r}_t)|$, $\mathbf{r}_t \in \mathcal{T}$, $t = 1, ..., T$) is reported in Fig. 3 (left column) for different *SNR*s. It turns out that the *NF* estimation error is always very small and upper-bounded to $\max_{\mathbf{r}_t \in \mathcal{T}} \left|\Delta \widetilde{E}(\mathbf{r}_t)\right|\Big|_{SNR=20\,[dB]} = -27.2$ [dB] [Fig. 3(*g*)]. For comparisons, the maps yielded with the *OMP*-based implementation [20] are reported (Fig. 3 - right column), as well, to pictorially underline the better performance of the *BCS* approach that are quantitatively confirmed by the plots of the corresponding errors (Fig. 4). As a matter of fact, the *BCS* is more robust to the noise and it remarkably reduces the *NF* error especially in the worst-case conditions [e.g., $SNR = 20$ [dB] - $\frac{\Xi|_{SNR=20\,[dB]}^{BCS}}{\Xi|_{SNR=20\,[dB]}^{OMP}} \approx -21$ [dB] and Fig. 3(*g*) vs. Fig. 3(*h*)]. Independently on the *SNR*, the *BCS* provides more sparse solutions than the *OMP* (i.e., $\left\|\widetilde{\underline{w}}^{BCS}\right\|_0 < \left\|\widetilde{\underline{w}}^{OMP}\right\|_0$) and, unlike this latter, the retrieved non-null entries of $\widetilde{\underline{w}}^{BCS}$ when $SNR > 20$ [dB] are in correspondence with the four redundant[1] basis vectors associated to the uncertainty factors affecting the actual *AUT* and highlighted with the vertical grey bar in Fig. 5. As a matter of fact, $\left|\widetilde{w}_b^{BCS}\right| > 0$ when $b = 5$ (i.e., the first column of $\underline{\underline{A}}$ linked to a variation of $|\zeta_3|$), $b = 25$, and $b = 26$ (i.e., the basis vectors corresponding to $\angle \zeta_3$) as shown in Figs 5(*a*)-5(*c*). This points out that the *BCS* does not only improve the estimation accuracy of the *NF* distribution in $\mathcal{T}$ of the *OMP*, but it also generally provides the information on which defect/anomaly is deviating the *AUT* pattern from the ideal one.

To investigate the impact of the *NF* prediction on the *FF* characterization of the *AUT*, let us analyze the error maps of the mismatch between the actual, $P(u, v) = \mathcal{F}\{E(\mathbf{r})\}$, and the retrieved, $\widetilde{P}(u, v) = \{\widetilde{E}(\mathbf{r})\}$, power patterns, $\left|\Delta \widetilde{P}(u, v)\right| \triangleq \left|P(u, v) - \widetilde{P}(u, v)\right|$, $u = \sin\theta \cos\phi$ and $v = \sin\theta \sin\phi$ being the direction cosines [$(u^2 + v^2) \leq 1$], while $\mathcal{F}\{.\}$ stands for the *NF-FF* operator. As a representative example, Figures 6(*a*)-6(*b*) refer to the case with $SNR = 20$ [dB]. As expected, the error of the *BCS* is smaller than that the *OMP* as a direct consequence of the more accurate *NF* reconstruction [Fig. 3(*g*) vs. Fig. 3(*h*) $\Rightarrow$ Fig. 6(*a*) vs. Fig. 6(*b*)]. Quantitatively, the maximum *FF* deviation $\left|\Delta \widetilde{P}(u, v)\right|_{\max}$ ($\left|\Delta \widetilde{P}(u, v)\right|_{\max} \triangleq \max_{(u^2+v^2) \leq 1} \left|\Delta \widetilde{P}(u, v)\right|$) is equal to $\left.\left|\Delta \widetilde{P}(u, v)\right|_{\max}^{BCS} = -34.2$ [dB] [Fig. 6(*a*)] and $\left.\left|\Delta \widetilde{P}(u, v)\right|_{\max}^{OMP} = -8.8$ [dB] [Fig. 6(*b*)], respectively. Such outcomes are further confirmed by the comparison between the actual and the retrieved power patterns along the most critical cut (i.e., $u = 0$) in Fig. 6(*c*). Indeed, the *OMP* curve clearly shows larger and non-negligible distortions of the pattern outside the main lobe region.

---

[1]Since the basis set is overcomplete, a *NF* field prediction needs at least one component for each uncertainty factor. In this example, there are two non-null entries for each of the $C = 2$ uncertainty factors. Therefore, one entry for each factor is mandatory.

Finally, let us investigate on the effectiveness of the proposed method in mitigating the "truncation error". Towards this end, the side of the measurement plane $\Psi_0$ has been shrunk to $L_{\Psi_0} = 12$ [$\lambda$] [Fig. 7(*a*)]. Regardless of the reduction of the probing area (i.e., $\frac{\Psi_0}{\Psi} = 0.36$) and the noise level ($SNR = 20$ [dB]), the *BCS* is very accurate in estimating the *NF* field distribution [Fig. 7(*c*)] and it clearly overcomes the *OMP* [Fig. 7(*c*) vs. Fig. 7(*e*)] as pointed out by the corresponding values of the integral error (i.e., $\Xi|_{\Psi_0=12[\lambda]}^{BCS} = -25.1$ [dB] vs. $\Xi|_{\Psi_0=12[\lambda]}^{OMP} = -4.5$ [dB]). Similar conclusions can be drawn also in the very challenging case of $L_{\Psi_0} = 8$ [$\lambda$] (i.e., $\frac{\Psi_0}{\Psi} = 0.16$) with errors equal to $\Xi|_{\Psi_0=8[\lambda]}^{BCS} = -22.8$ [dB] [Fig. 7(*d*)] and $\Xi|_{\Psi_0=8[\lambda]}^{OMP} = -2.7$ [dB]. For completeness, the ($u = 0$)-cut of the corresponding *FF* patterns are shown in Fig. 8, where the transformed pattern from the actual truncated *NF* data, $P_0(u, v) = \mathcal{F}\left\{E(\mathbf{r})|_{\mathbf{r} \in \Psi_0}\right\}$, is reported as well, to better highlight the reconstruction capabilities of the *BCS* solution technique and its superior capability of mitigating the truncation error.

# 4 Conclusions

A novel antenna characterization technique, which leverages on the *MebD* paradigm and a suitable implementation of the *BCS*, has been proposed to faithfully recover the *NF* field distribution generated by an *AUT* starting from a limited set of measurements. More specifically, the *NF* reconstruction problem has been reformulated within a probabilistic sparsity-promoting framework to bypass the cumbersome check of the *RIP* compliancy of the measurement operator. An analysis of the dependence of the *BCS* performance on its calibration setup has been carried out to derive some practical guidelines for its reliable application. The potentialities and limitations of the proposed method have been assessed in a comparative numerical validation under different operative conditions. The main outcomes from the numerical analyses are:

- the proposed *BCS*-based method faithfully retrieves the *NF* behavior of the *AUT* with a remarkable robustness to the noise. This implies a reliable characterization of the *FF* radiation pattern of the *AUT* in the whole visible range;

- unlike other state-of-the-art *CS* alternative methods, it not only yields an accurate *NF/FF* prediction of the radiated field, but it also provides information on which uncertainties are causing the deviations of the *AUT* pattern from the gold one;

- it exhibits excellent performance in mitigating the truncation error caused by limited-extension probing area.

Finally, it is worth pointing out that the proposed method is general and it is extendable to any type of antenna or field distribution. Moreover, it can handle arbitrarily-shaped non-planar characterization setups and different measurement topologies (e.g., spiral scanning [13]).

# Acknowledgements

This work has been partially supported by the Italian Ministry of Education, University, and Research within the Program PRIN 2017 (CUP: E64I19002530001) for the Project "EMvisioning" (Grant no. 2017HZJXSZ), within the Program "Smart cities and communities and Social Innovation" (CUP: E44G14000040008) for the Project "SMARTOUR" (Grant no. SCN_00166), within the Program PON R&I 2014-2020 for the Project "MITIGO" (Grant no. ARS01_00964), and under Grant "Dipartimenti di Eccellenza (2018-2022)". Moreover, it benefited from the networking activities carried out within the Project "SPEED" (Grant No. 61721001) funded by National Science Foundation of China under the Chang-Jiang Visiting Professorship Program. A. Massa wishes to thank E. Vico for her never-ending inspiration, support, guidance, and help.

# FIGURE CAPTIONS

- **Figure 1.** Pictorial sketch of the *NF* antenna characterization scenario.

- **Figure 2.** *BCS Calibration* ($H = 7$ [$\lambda$], $L_\Psi = L_{\Psi_0} = 20$ [$\lambda$], $M = 25$, $T = 1681$, $N = 60$, $|\zeta_3| = 0.45$, $\angle\zeta_3 = \frac{\pi}{3}$ [rad], $SNR \in [20, 50]$ [dB]) - Behavior of the *NF* integral error, $\Xi$, as a function of the *BCS* parameter $\eta_0$.

- **Figure 3.** *Numerical Validation* ($H = 7$ [$\lambda$], $L_\Psi = L_{\Psi_0} = 20$ [$\lambda$], $M = 25$, $T = 1681$, $N = 60$, $|\zeta_3| = 0.45$, $\angle\zeta_3 = \frac{\pi}{3}$ [rad]) - *NF* error maps $\left|\Delta\widetilde{E}(\mathbf{r})\right|$ yielded by the (*a*)(*c*)(*e*)(*g*) *BCS* and the (*b*)(*d*)(*f*)(*h*) *OMP* when processing noisy data with (*a*)(*b*) $SNR = 50$ [dB], (*c*)(*d*) $SNR = 40$ [dB], (*e*)(*f*) $SNR = 30$ [dB], and (*g*)(*h*) $SNR = 20$ [dB].

- **Figure 4.** *Numerical Validation* ($H = 7$ [$\lambda$], $L_\Psi = L_{\Psi_0} = 20$ [$\lambda$], $M = 25$, $T = 1681$, $N = 60$, $|\zeta_3| = 0.45$, $\angle\zeta_3 = \frac{\pi}{3}$ [rad], $SNR \in [20, 50]$ [dB]) - Behavior of the *NF* integral error, $\Xi$, as a function of the *SNR*.

- **Figure 5.** *Numerical Validation* ($H = 7$ [$\lambda$], $L_\Psi = L_{\Psi_0} = 20$ [$\lambda$], $M = 25$, $T = 1681$, $N = 60$, $|\zeta_3| = 0.45$, $\angle\zeta_3 = \frac{\pi}{3}$ [rad]) - Magnitude of the retrieved expansion coefficients, $|\widetilde{w}_b|$, $b = 1, ..., B$, outputted by the *BCS* and the *OMP* methods when processing noisy data with (*a*) $SNR = 50$ [dB], (*b*) $SNR = 40$ [dB], (*c*) $SNR = 30$ [dB], and (*d*) $SNR = 20$ [dB].

- **Figure 6.** *Numerical Validation* ($H = 7$ [$\lambda$], $L_\Psi = L_{\Psi_0} = 20$ [$\lambda$], $M = 25$, $T = 1681$, $N = 60$, $|\zeta_3| = 0.45$, $\angle\zeta_3 = \frac{\pi}{3}$ [rad], $SNR = 20$ [dB]) - *FF* error maps, $\left|\Delta\widetilde{P}(u, v)\right|$, yielded by (*a*) the *BCS* and (*b*) the *OMP* methods. Estimated power patterns along the ($u = 0$)-cut (*c*).

- **Figure 7.** *Numerical Validation* ($H = 7$ [$\lambda$], $L_\Psi = 20$ [$\lambda$], $M = 25$, $T = 1681$, $N = 60$, $|\zeta_3| = 0.45$, $\angle\zeta_3 = \frac{\pi}{3}$ [rad], $SNR = 20$ [dB]) - (*a*)(*b*) Measurement setup and (*c*)-(*f*) *NF* error maps yielded by the (*c*)(*d*) *BCS* and the (*e*)(*f*) *OMP* methods when considering a truncated *NF* probing region of side (*a*)(*c*)(*e*) $L_{\Psi_0} = 12$ [$\lambda$] and (*b*)(*d*)(*f*) $L_{\Psi_0} = 8$ [$\lambda$].

- **Figure 8.** *Numerical Validation* ($H = 7$ [$\lambda$], $L_\Psi = 20$ [$\lambda$], $M = 25$, $T = 1681$, $N = 60$,

$|\zeta_3| = 0.45$, $\angle \zeta_3 = \frac{\pi}{3}$ [rad], $SNR = 20$ [dB]) - Actual, $P(u, v)$, truncated, $P_0(u, v)$, and estimated, $\widetilde{P}(u, v)$, power patterns along the $(u = 0)$-cut when (*a*) $L_{\Psi_0} = 12$ [$\lambda$] and (*b*) $L_{\Psi_0} = 8$ [$\lambda$].

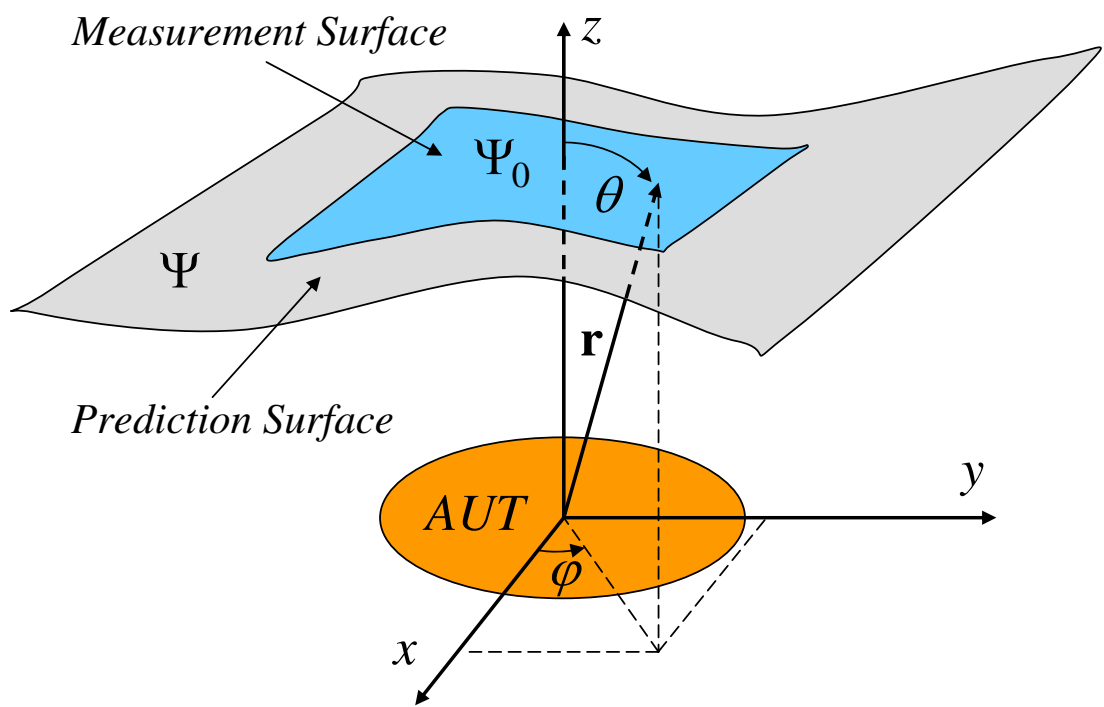

**Fig. 1** - M. Salucci *et al.*, "A Bayesian Compressive Sensing Approach ..."

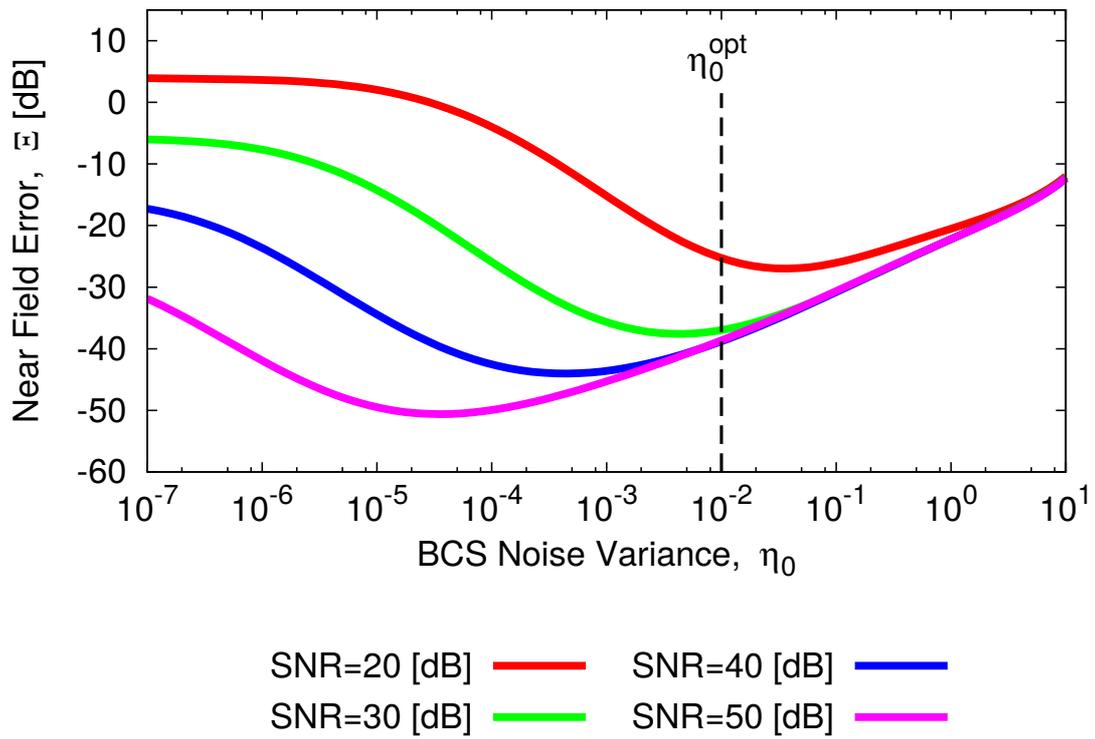

**Fig. 2 - M. Salucci *et al.*,** "A Bayesian Compressive Sensing Approach ..."

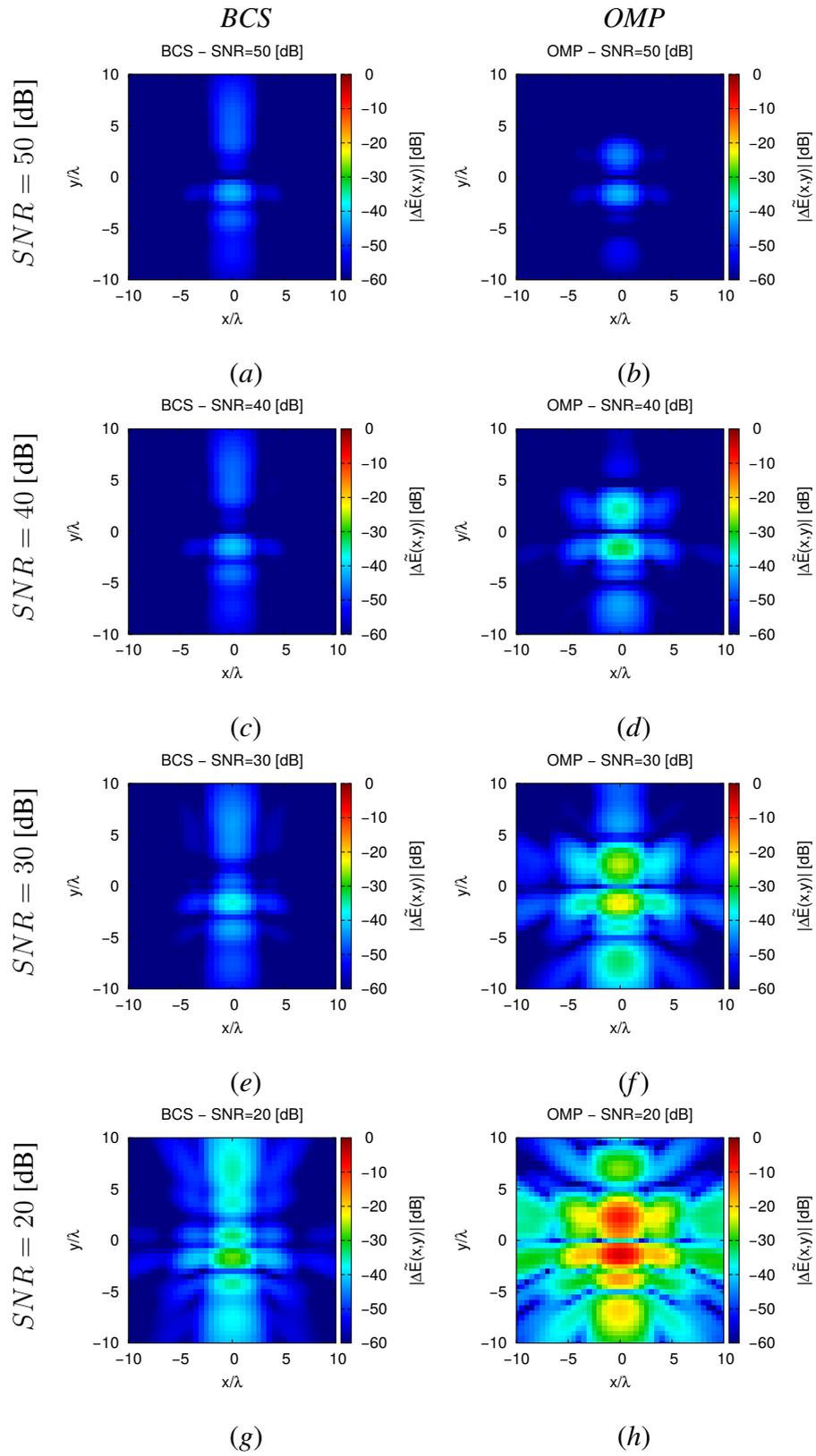

**Fig. 3** - M. Salucci *et al.*, "A Bayesian Compressive Sensing Approach ..."

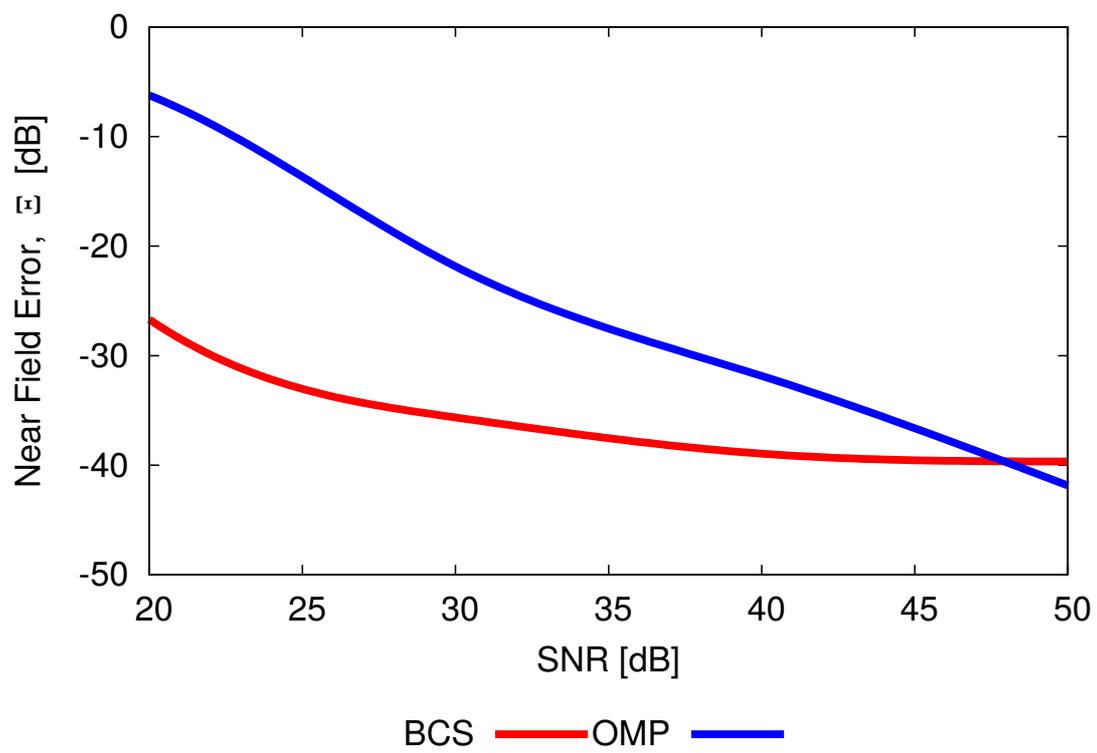

**Fig. 4 - M. Salucci *et al.*,** "A Bayesian Compressive Sensing Approach ..."

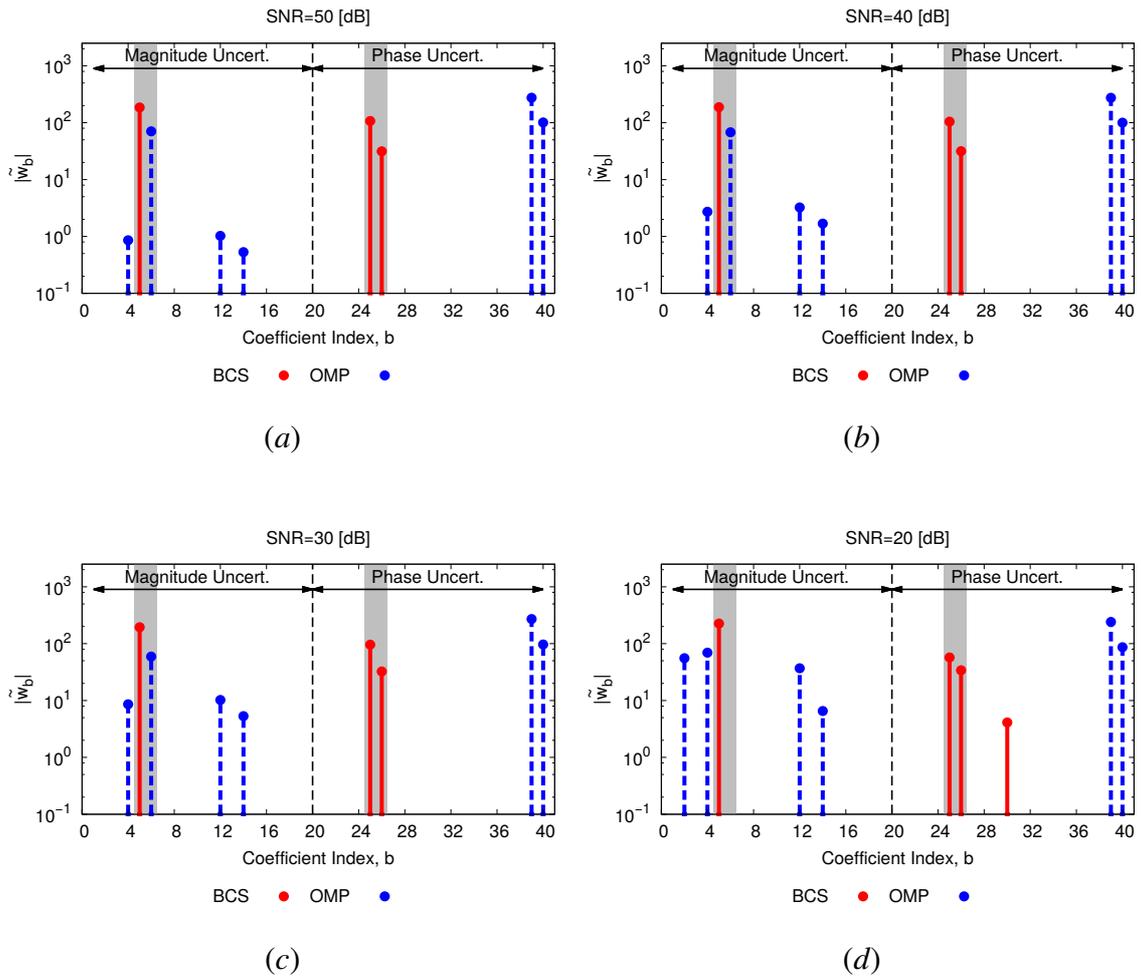

**Fig. 5 - M. Salucci *et al.*,** "A Bayesian Compressive Sensing Approach ..."

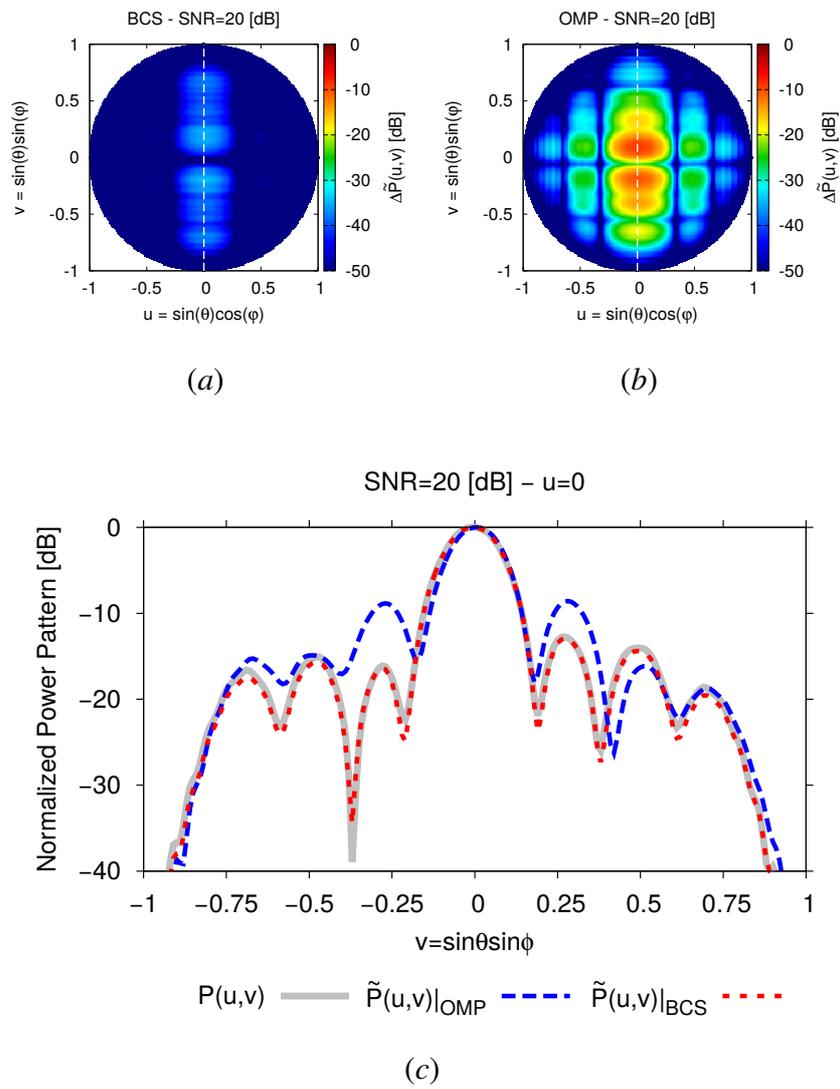

**Fig. 6** - M. Salucci *et al.*, "A Bayesian Compressive Sensing Approach ..."

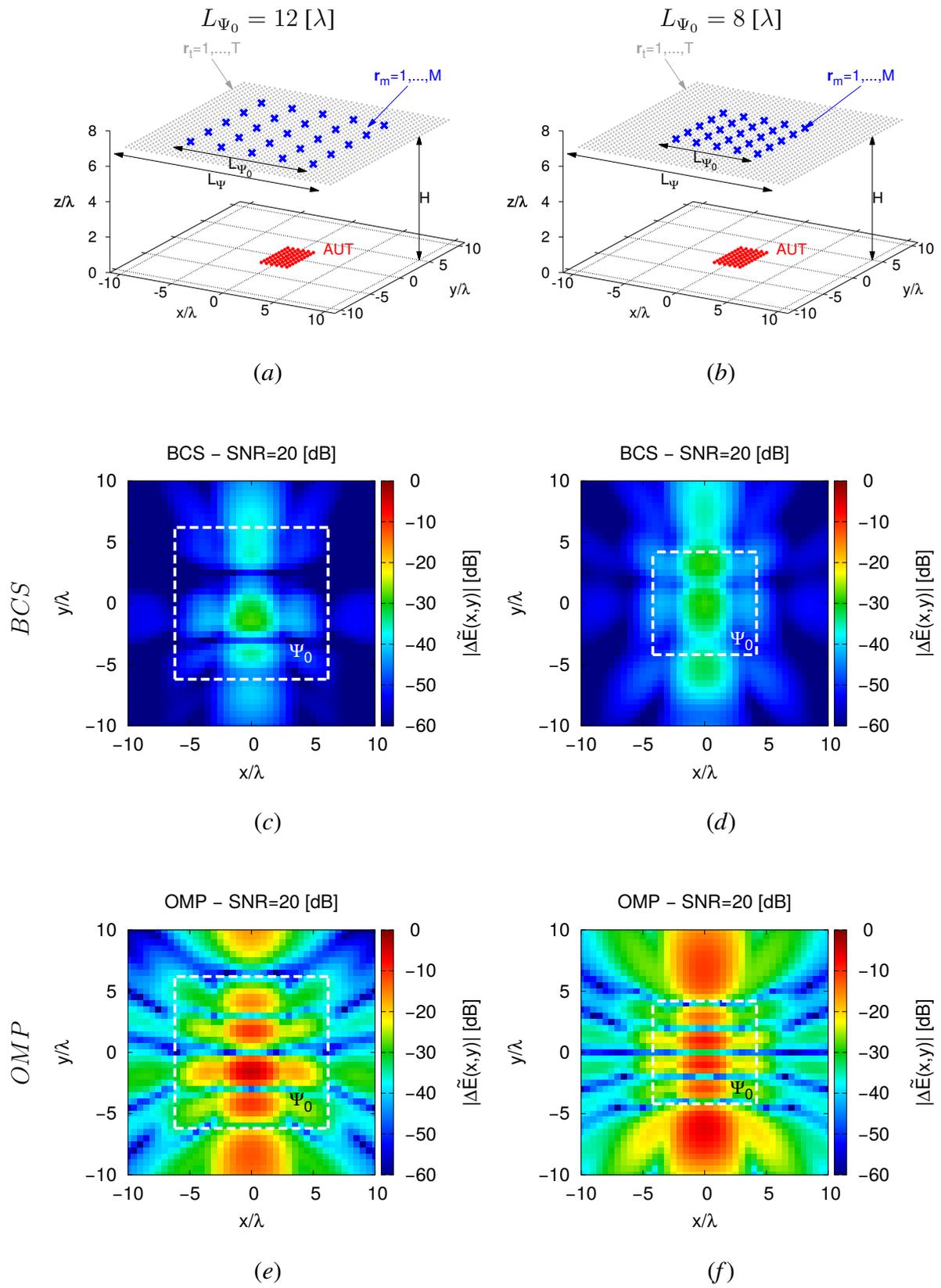

**Fig. 7** - M. Salucci *et al.*, "A Bayesian Compressive Sensing Approach ..."

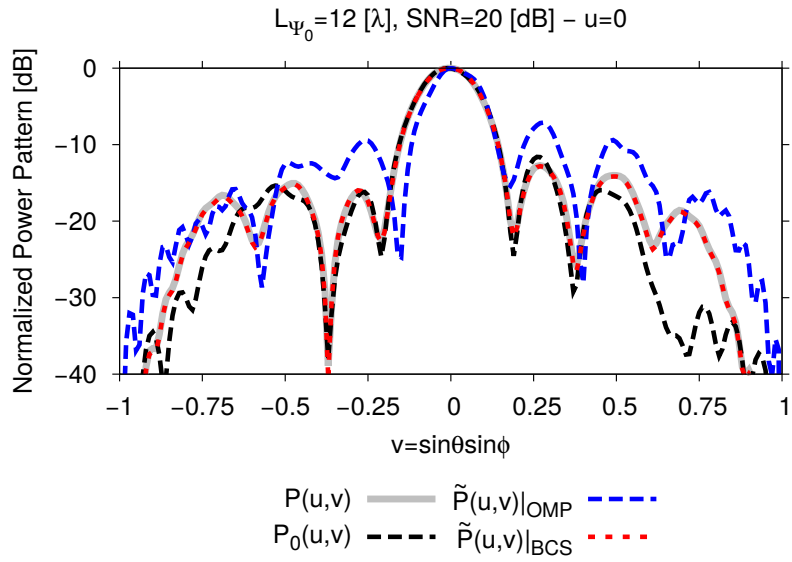

(a)

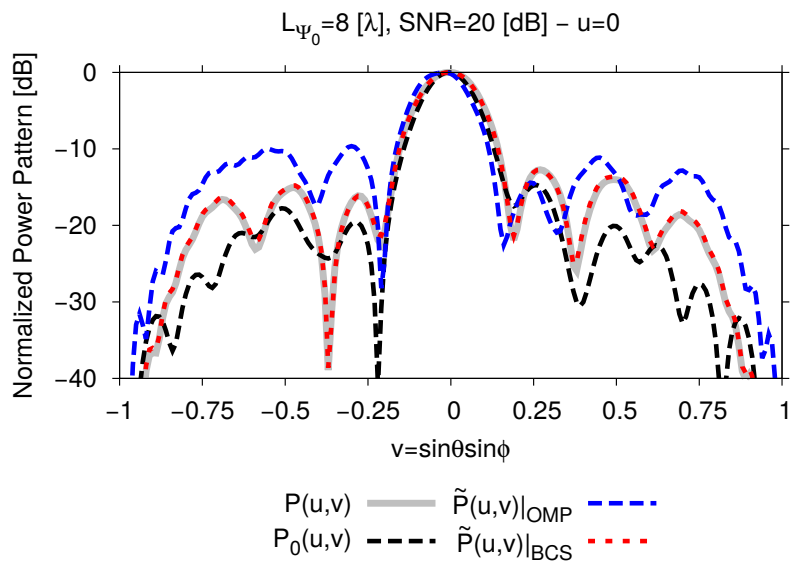

(b)

**Fig. 8** - M. Salucci *et al.*, "A Bayesian Compressive Sensing Approach ..."